# Mathematical Model of Gravitational and Electrostatic Forces

Alexei Krouglov

Email: alexkrouglov@gmail.com

ABSTRACT

Author presents mathematical model for acting-on-a-distance attractive and repulsive forces based on propagation of energy waves that produces Newton expression for gravitational and Coulomb expression for electrostatic forces. Model uses mathematical observation that difference between two inverse exponential functions of the distance asymptotically converges to function proportional to reciprocal of distance squared.

# 1. Introduction

Author described his *Dual Time-Space Model of Wave Propagation* in work [2]. He tries to express physical phenomena based only on concept of energy.

Model asserts that when at some point in space and in time local energy value is different from global energy level in surrounding area, energy disturbance is created at this point. Energy disturbance propagates both in time and in space as energy wave, and oscillates. Author used the model to describe both propagation of energy waves in physics [3] – [5] and fluctuations of demand and supply in economics [6], [7].

Author presented in [3] mathematical model for attractive and repulsive forces that demonstrated how gravitational and electrostatic forces are created on a distance based on instrument of *Dual Time-Space Model of Wave Propagation*. However paper [3] didn't explain one important point. It described forces acting on a distance as being proportional to inverse exponential function of distance. Lately author realized that difference between two inverse exponential functions of distance is asymptotically converging to function that is proportional to reciprocal of distance squared. Thus result provided by the model is in agreement with experimentally obtained formulas for forces acting on a distance such as gravitational and electrostatic forces.

In current paper author presents quantitative mathematical model for acting-on-a-distance static forces.

## 2. Model Methodology

As it was described in [2], each point in space and in time is characterized by two energy values – local energy value in point $U(x,t)$ and global energy level in neighboring area $\Phi(x,t)$. The difference between quantities $U(x,t)$ and $\Phi(x,t)$ creates energy disturbance, which propagates both in time and in space according to ordinary differential equations presented below.

Differential equations in time domain are,

$$\frac{dP(x_0,t)}{dt} = -\lambda_t (U(x_0,t) - \Phi(x_0,t)) \tag{1}$$

$$\frac{d^2U(x_0,t)}{dt^2} = \mu_t \frac{dP(x_0,t)}{dt} \tag{2}$$

$$\frac{d\Phi(x_0,t)}{dt} = -\nu_t \frac{dP(x_0,t)}{dt} \tag{3}$$

where $x_0 \in \Re$, $\lambda_t$, $\mu_t$, $\nu_t \geq 0$ are constants, and $P(x_0,t)$ is *dual variable* evaluated in point $x_0$ at time $t$.

Briefly speaking, meaning of equations (1) – (3) is as follows. Second derivative of energy value $U(x,t)$ with respect to time is inversely proportional to energy disturbance $U(x,t) - \Phi(x,t)$. First derivative of energy level $\Phi(x,t)$ with respect to time is directly proportional to energy disturbance $U(x,t) - \Phi(x,t)$.

We can rewrite equations (1) – (3) to describe dynamics of changes for energy disturbance $E(x_0,t) = U(x_0,t) - \Phi(x_0,t)$ in time,

$$\frac{d^2 E(x_0,t)}{dt^2} + \nu_t \lambda_t \frac{dE(x_0,t)}{dt} + \mu_t \lambda_t E(x_0,t) = 0 \qquad (4)$$

Equation (4) is an ordinary differential equation of the second order, and can be resolved by regular methods (for example, see [1]). Note that value $E(x_0,t) \to 0$ when $t \to +\infty$ for all $x_0$.

Differential equations in space domain are,

$$\frac{dP(x,t_0)}{dx} = -\lambda_x (U(x,t_0) - \Phi(x,t_0)) \qquad (5)$$

$$\frac{d^2 U(x,t_0)}{dx^2} = \mu_x \frac{dP(x,t_0)}{dx} \qquad (6)$$

$$\frac{d\Phi(x,t_0)}{dx} = -\nu_x \frac{dP(x,t_0)}{dx} \qquad (7)$$

where $t_0 \in \Re^+$, $\lambda_x, \mu_x, \nu_x \geq 0$ are constants.

Similarly as above, meaning of equations (5) – (7) is as follows. Second derivative of energy value $U(x,t)$ with respect to the distance in space is inversely proportional to energy disturbance $E(x,t)$. First derivative of energy level $\Phi(x,t)$ with respect to the distance in space is directly proportional to energy disturbance $E(x,t)$.

We can rewrite equations (5) – (7) to describe dynamics of changes for energy disturbance $E(x,t_0)$ in space,

$$\frac{d^2 E(x,t_0)}{dx^2} + \nu_x \lambda_x \frac{dE(x,t_0)}{dx} + \mu_x \lambda_x E(x,t_0) = 0 \qquad (8)$$

Equation (8) is also an ordinary differential equation of the second order. Here again value $E(x,t_0) \to 0$ when $x \to +\infty$ for all $t_0 > 0$.

One of primary characteristics of current model is that it uses three variables to describe dynamics of wave propagation. Nonetheless present paper is devoted to static phenomena, particularly to formation of static forces acting on a distance between solid bodies. Dynamics of wave propagation in space will involve a special study.

## 3. Transformation of Global Energy Level near Solid Body

Author considers in current paper the static situation $t \equiv t_0$. It means that energy disturbance is not propagated in time. Here we deal with solid bodies – it means from the model's point of view that energy disturbance is not propagated in space with regard to local energy values $U(x,t)$. Thus we only consider transformation of global energy level $\Phi(x,t)$.

Mathematically speaking, static situation means that equations (1) – (3) have null coefficients $\lambda_t, \mu_t, \nu_t = 0$. Solid body means null coefficient $\mu_x = 0$ in equations (5) – (7) (other coefficients in equations (5) – (7) are nonnegative $\lambda_x, \nu_x \geq 0$). Therefore equations (5) and (7) can be combined into the following one,

$$\frac{d\Phi(x,t)}{dx} = -\nu_x \lambda_x (\Phi(x,t) - U(x,t)) \tag{9}$$

Since we are dealing with solid body, local energy value $U(x,t)$ can't propagate behind body's boundaries $x_0 \in \Re$.

The situation is illustrated on **Figure 1**, which shows transformation of energy level in space near boundary of solid body.

**Figure 1a** shows situation when no energy disturbance is present and local energy values in space match global energy level in surrounding area. **Figure 1b** displays situation on the boundary of solid body where local energy value abruptly changes its magnitude. That change creates energy disturbance, which transforms global energy level behind the boundaries of solid body.

To calculate transformation of global energy level one has to solve equation (9) for $x_1 > x_0$. We have following initial conditions at the body's boundary $x_0 \in \Re$ for $\forall t > 0$, $U(x_0, t) \equiv \Phi(x_0, t) \equiv U_0$.

We assume that $U(x_1, t) \equiv U_1$ is constant for $x_1 > x_0$ and $\forall t > 0$, and equation (9) for global energy level $\Phi(x, t)$ can produce following equation for energy disturbance $E(x, t) = U_1 - \Phi(x, t)$,

$$\frac{dE(x,t)}{dx} = -\nu_x \lambda_x \, E(x,t) \tag{10}$$

where $x > x_0$, $\forall t > 0$, and $E(x_0, t) = U(x_1, t) - \Phi(x_0, t) = U_1 - U_0 = -\Delta U$.

Thus solution of equation (10) is as follows for $x > x_0$,

$$E(x,t) = E_0 \exp\{-\nu_x \lambda_x \, (x - x_0)\} \tag{11}$$

where $E_0 = -\Delta U$.

Therefore solution of equation (9) is for $x > x_0$,

$$\Phi(x,t) = (U_0 - U_1)\exp\{-\nu_x \lambda_x (x - x_0)\} + U_1 \qquad (12)$$

since $\Delta U = -E_0 = U_0 - U_1$.

**Figure 1c** illustrates how global energy level is transformed near the boundary of solid body according to equation (12).

## 4. Forces near Solid Body at Rest

As we know, concept of physical forces is related with ideas of work and distance, and particularly – multiplication of average force applied to body by distance equals to the work, where work equals to change of body's energy.

Following to discussion above, when a solid body has energy disturbance, it creates energy disbalance in surrounding area. If energy disbalance occurs on the boundary of solid body, the body is trying to move to another position in order to reduce this energy disbalance. Thus energy disturbance creates physical force. The magnitude of that force is proportional to changes of absolute value of energy disturbance with respect to distance in space. Force points to direction of decrease of absolute value of energy disturbance. Next definition of force is used in the model.

**Definition.** *Instantaneous change of absolute value of energy disturbance at the boundary of solid body* $\frac{d|E(x_0,t)|}{dx}$ *in point* $x_0$ *with respect to distance in space is called force*

$F(x_0, t)$ *applied to solid body in point* $x_0$ *and directed to the decrease of absolute value of energy disturbance,*

$$F(x_0, t) \equiv -k \frac{d|E(x_0, t)|}{dx} = -k \frac{d|U(x_0, t) - \Phi(x_0, t)|}{dx} \tag{13}$$

where $k > 0$ is coefficient of proportionality.

Let's look how this definition of force works for solid body that is placed at rest.

The situation with solid body at rest is illustrated on **Figure 2**.

**Figure 2a** displays local energy values $U(x,t)$ and global energy level $\Phi(x,t)$ near solid body. **Figure 2b** presents distribution of energy disturbance $E(x,t)$ near solid body. **Figure 2c** examines adjusted distribution of energy disturbance $E(x,t)$ near solid body, which takes into account transformation of global energy level on each side of the solid body's border (this correction of global energy level is not further used in the paper).

Thus if points in space $x_0 \in \Re$ and $x_2 \in \Re$, where $x_0 > x_2$, are respectively right and left boundaries of solid body, transformation of global energy level $\Phi(x,t)$ behind right and left boundaries of solid body is described as,

$$\Phi(x,t) = \begin{cases} \Delta U \exp\{-\nu_x \lambda_x (x - x_0)\} + U_1 & if \quad x > x_0 \\ \Delta U \exp\{\nu_x \lambda_x (x - x_2)\} + U_1 & if \quad x < x_2 \end{cases} \tag{14}$$

Then forces $F(x_0, t)$ and $F(x_2, t)$ acting on solid body at its boundaries are as follows (function $|E(x,t)| = |U_1 - \Phi(x,t)|$ is decreasing to the right of body's boundary $x > x_0$, and is increasing to the left of body's boundary $x < x_2$),

$$F(x_0,t) = -\lim_{\substack{x \to x_0 \\ x > x_0}} k \frac{|E(x,t)| - |E(x_0,t)|}{x - x_0} = -\lim_{\substack{x \to x_0 \\ x > x_0}} k \frac{\Phi(x,t) - \Phi(x_0,t)}{x - x_0} = -k \frac{d\Phi(x_0,t)}{dx} = \nu_x \lambda_x k \Delta U \quad (15)$$

$$F(x_2,t) = -\lim_{\substack{x \to x_2 \\ x < x_2}} k \frac{|E(x,t)| - |E(x_2,t)|}{x - x_2} = -\lim_{\substack{x \to x_2 \\ x < x_2}} k \frac{\Phi(x,t) - \Phi(x_2,t)}{x - x_2} = -k \frac{d\Phi(x_2,t)}{dx} = -\nu_x \lambda_x k \Delta U \quad (16)$$

Since we assumed above that solid body is at rest (i.e. forces at body's boundaries are in balance with each other), it follows,

$$F(x_0,t) + F(x_2,t) = 0 \quad (17)$$

## 5. Forces around Two Solid Bodies

In this section we study impact of energy disturbance on the boundaries of one solid body that affects energy disturbance on the boundaries of another solid body (as displayed on **Figure 3**). We consider three points in space $x_0 \in \Re$, $x_1 \in \Re$, and $x_3 \in \Re$ where $x_3 > x_1 > x_0$. Here $x_0$ is right boundary of first solid body, $x_1$ and $x_3$ are left and right boundaries respectively of second solid body (see **Figure 3a**).

If we denote as $U_2$ – local energy value for second body and as $\Delta U' = U_2 - U_1$ –jump of energy value on the boundaries of second body, then reverse forces, which hold second body at rest, are as before,

$$F(x_1, t) = -\nu_x \lambda_x k \, \Delta U' \tag{18}$$

$$F(x_3, t) = \nu_x \lambda_x k \, \Delta U' \tag{19}$$

$$F(x_1, t) + F(x_3, t) = 0 \tag{20}$$

When we consider resultant energy disturbance near second body, we have to consider not only transformation of global energy level to the left of point $x_1$ – left boundary of second body and to the right of point $x_3$ – right boundary of second body but transformation of global energy level to the right of point $x_0$ – right boundary of first body as well. Dual effect of transformations of global energy level near second body is reflected by subtracting from global energy level transformed by boundaries of second body $x_1$ and $x_3$ the global energy level transformed by right boundary $x_0$ of first body. We may say that we adjust global energy level near solid body by subtracting from it the impact induced by other solid bodies in surrounding area.

**Figure 3b** shows transformations of global energy level $\Phi_1(x,t)$, $\Phi_2(x,t)$, and $\Phi_3(x,t)$ caused by right boundary $x_0$ of first body, left boundary $x_1$ of second body, and right boundary $x_3$ of second body respectively.

**Figure 3c** displays distributions of energy disturbances $E_1(x,t)$, $E_2(x,t)$, and $E_3(x,t)$ caused by right boundary $x_0$ of first body, left boundary $x_1$ and right boundary $x_3$ of second body respectively if the energy disturbances were separate.

Now we want to move from separate energy disturbances $E_1(x,t)$, $E_2(x,t)$, and $E_3(x,t)$ to resultant energy disturbance $\hat{E}(x,t)$ at the boundaries of second body $x_1$ and $x_3$. We assume that energy disturbances $E_2(x,t)$ and $E_3(x,t)$ in points $x_1$ and $x_3$ are decreased by energy disturbance $E_1(x,t)$ propagated from point $x_0$ of first body. Thus energy disturbances $E_2(x_1,t)$ and $E_3(x_3,t)$ in points $x_1$ and $x_3$ are decreased by values $E_1(x_1,t)$ and $E_1(x_3,t)$ respectively. It gives us resultant energy disturbances $\hat{E}(x_1,t)$ and $\hat{E}(x_3,t)$ in points $x_1$ and $x_3$, and creates forces $\hat{F}(x_1,t)$ and $\hat{F}(x_3,t)$ acting on boundaries of second body.

Discussion above can be summarized by following assertion.

**Assumption.** *Quantity of energy disturbance* $E(x_0,t)$ *in point* $x_0$ *– boundary of solid body is decreased by sum of energy disturbances* $\sum_i E_i(x_0,t)$ *propagated to point* $x_0$ *from other solid bodies in vicinity,*

$$\hat{E}(x_0,t) \equiv E(x_0,t) - \sum_i E_i(x_0,t) \qquad (21)$$

We may look at equation (21) from a new point of view – energy disturbances at the boundaries of solid body are summed with energy disturbances propagated to these boundaries from reflected images of other solid bodies in vicinity (*reflected image* of solid body has energy disturbance of opposite sign).

**Figure 3d** presents distribution of resultant energy disturbance $\hat{E}(x,t)$ near the boundaries of second body in points $x_1$ and $x_3$, and shows forces $\hat{F}(x,t)$ acting on second body in points $x_1$ and $x_3$.

To calculate forces $\hat{F}(x_1,t)$ and $\hat{F}(x_3,t)$ in points $x_1$ and $x_3$ we denote as $\Delta U_1$ and as $\Delta U_3$ – the absolute values of resultant energy disturbance $\hat{E}(x,t)$ in points $x_1$ and $x_3$. Then forces acting on the second body in these points are,

$$\hat{F}(x_1,t) = -\nu_x \lambda_x k \, \Delta U_1 \tag{22}$$

$$\hat{F}(x_3,t) = \nu_x \lambda_x k \, \Delta U_3 \tag{23}$$

where $\Delta U_1 = \left|\hat{E}(x_1,t)\right|$ and $\Delta U_3 = \left|\hat{E}(x_3,t)\right|$.

Consequently resultant force acting on second body is as follows,

$$\hat{F}(x_1,t) + \hat{F}(x_3,t) = \nu_x \lambda_x k \left(\Delta U_3 - \Delta U_1\right) \tag{24}$$

On the other hand, absolute values of resultant energy disturbances $\Delta U_1$ and $\Delta U_3$ at the boundaries of second body in points $x_1$ and $x_3$ are as follows,

$$\Delta U_1 = \left|E_2(x_1,t) - E_1(x_1,t)\right| = \left|\Phi_2(x_1,t) - \Phi_1(x_1,t)\right| = \left|\Delta U' - \Delta U \exp\{-\nu_x \lambda_x (x_1 - x_0)\}\right| \tag{25}$$

$$\Delta U_3 = \left|E_3(x_3,t) - E_1(x_3,t)\right| = \left|\Phi_3(x_3,t) - \Phi_1(x_3,t)\right| = \left|\Delta U' - \Delta U \exp\{-\nu_x \lambda_x (x_3 - x_0)\}\right| \tag{26}$$

Let us assume for simplicity further that $\Delta U' \geq \Delta U \exp\{-\nu_x \lambda_x (x_1 - x_0)\}$ and $\Delta U' \geq \Delta U \exp\{-\nu_x \lambda_x (x_3 - x_0)\}$. Then it follows from equations (25) and (26),

$$\Delta U_1 = \Delta U' - \Delta U \exp\{-\nu_x \lambda_x (x_1 - x_0)\} \tag{27}$$

$$\Delta U_3 = \Delta U' - \Delta U \exp\{-\nu_x \lambda_x (x_3 - x_0)\} \tag{28}$$

and it follows from equation (24),

$$\hat{F}(x_1, t) + \hat{F}(x_3, t) = \nu_x \lambda_x \, k \, \Delta U \left(\exp\{-\nu_x \lambda_x (x_1 - x_0)\} - \exp\{-\nu_x \lambda_x (x_3 - x_0)\}\right) \tag{29}$$

Therefore resultant force $\hat{F}(x_1, t) + \hat{F}(x_3, t) > 0$ when $\Delta U > 0$ and $x_3 > x_1 > x_0$. It means that resultant force $\hat{F}(x_1, t) + \hat{F}(x_3, t)$ is directed to the right of point $x_3$ of second body, i.e. second body is repulsed from first body in this situation.

Let us estimate resultant force on second body $\hat{F}_2(t) = \hat{F}(x_1, t) + \hat{F}(x_3, t)$ presented by equation (29).

Exponential function can be represented in series expansion,

$$\exp\{\tau\} = 1 + \frac{\tau}{1} + \frac{\tau^2}{2} + \frac{\tau^3}{6} + \ldots \tag{30}$$

We approximate the exponential function by the truncated series expansion,

$$\exp\{\tau\} \cong 1 + \frac{\tau}{1}$$

and

$$\exp\{-\tau\} \cong \frac{1}{1+\tau}$$

Then for $\tau_2 > \tau_1 > 0$,

$$\exp\{-\tau_1\} - \exp\{-\tau_2\} \approx \frac{1}{1+\tau_1} - \frac{1}{1+\tau_1+\delta} = \frac{1+\tau_1+\delta-1-\tau_1}{(1+\tau_1)(1+\tau_1+\delta)} = \frac{\delta}{(1+\tau_1)(1+\tau_1+\delta)}$$

$$= \frac{\delta}{\tau_1(\tau_1+\delta)\left(1+\frac{1}{\tau_1}\right)\left(1+\frac{1}{\tau_1+\delta}\right)} \approx \frac{\delta}{\tau_1(\tau_1+\delta)} \approx \frac{\delta}{\tau_1^2} \qquad (31)$$

where $\delta = \tau_2 - \tau_1$ and $\tau_1 >> \delta$.

As a result, by combining equations (29) and (31) we obtain next approximation for resultant force $\hat{F}_2(t)$ acting on second body,

$$\hat{F}_2(t) \cong k\,\Delta U\,\frac{r}{R^2} \qquad (32)$$

where $R = x_1 - x_0$, $r = x_3 - x_1$, and $R >> r$.

Thus we have established *Inverse Square Law* for resultant force $\hat{F}_2(t)$ acting on second body and given by equation (32).

Let us extend formula for resultant force on two- and three-dimensional spaces.

At first we refer to observation that resultant force acting on solid body is obtained by summarizing all forces acting on that body.

**Assumption.** *Resultant force acting on a solid body is equal to sum of all forces acting on the body,*

$$\hat{F}(t) \equiv \sum_i F_i(t) \tag{33}$$

Secondly, if we move from one-dimensional space that was considered so far to multi-dimensional spaces, we imagine plurality of lines starting perpendicularly to the surface of first solid body, and penetrating the surface of second solid body. One line produces forces on the boundaries of second body. Each resultant force is proportional to interval of the line that is contained between the boundaries of second body (as stated by equation (32)). To obtain total resultant force $\hat{F}_2(t)$ acting on second body we have to sum together all resultant forces acting at boundaries of the body. Total resultant force acting on second body is proportional to the sum of all such intervals that are located between the boundaries of second body.

Thirdly, when we consider solid bodies that are relatively departed from each other (i.e. $R >> r$ as stated above), the lines connecting two bodies are drawn in parallel to each other. Therefore for two-dimensional space sum of all parallel intervals that lie between boundaries of two-dimensional body equals to the body's internal area. Respectively for three-dimensional space sum of all parallel intervals that lie between boundaries of three-dimensional body equals to the body's internal volume.

The discussion above is illustrated on **Figure 4**.

**Figure 4a** shows forces, whose sum makes resultant $\hat{F}_2(t)$ force acting on two-dimensional second body. Similarly **Figure 4b** demonstrates forces, whose sum creates resultant force $\hat{F}_2(t)$ acting on three-dimensional second body.

Therefore in three-dimensional space resultant force $\hat{F}_2(t)$ acting on second solid body is expressed by following equation,

$$\hat{F}_2(t) \cong k\,\Delta U\,\frac{V}{R^2} \tag{34}$$

where $R$ is distance between first and second bodies, $V$ is second body's internal volume, $\Delta U$ is jump of energy value on the boundaries of first body, and $k > 0$ is coefficient of proportionality.

Since indexing on first and second bodies can be interchanged, we obtain pair of expressions for absolute values of $\hat{F}_1(t)$ and $\hat{F}_2(t)$ – resultant forces acting on first and second bodies respectively,

$$\left|\hat{F}_1(t)\right| \sim \Delta U_2\,\frac{V_1}{R^2} \tag{35}$$

$$\left|\hat{F}_2(t)\right| \sim \Delta U_1\,\frac{V_2}{R^2} \tag{36}$$

where $\Delta U_1$ and $\Delta U_2$ are jumps of energy values on boundaries of first and second bodies respectively, $V_1$ and $V_2$ are internal volumes of first and second bodies respectively, and $R$ is distance between first and second bodies.

At the same time, from Newton's third law resultant forces $\hat{F}_1(t)$ and $\hat{F}_2(t)$ acting on first and second bodies respectively should have the same absolute values, and point to opposite directions.

Therefore it takes place for $\left|\hat{F}(t)\right| = \left|\hat{F}_1(t)\right| = \left|\hat{F}_2(t)\right|$,

$$\left|\hat{F}(t)\right| \sim \frac{(\Delta U_2\,V_1)(\Delta U_1\,V_2)}{R^2} \tag{37}$$

or equivalently,

$$\left|\hat{F}(t)\right| \cong K\frac{(\Delta U_1 V_1)(\Delta U_2 V_2)}{R^2} \tag{38}$$

where $K > 0$ is coefficient of proportionality.

## 6. Electrostatic Forces

To get Coulomb's formula for electrostatic forces we consider two separate cases – electrostatic forces for repulsion and attraction.

***a) Repulsive Electrostatic Forces***

At first, I consider for diversity one-dimensional space with two solid bodies, each with negative jump of energy values $\Delta U = U_0 - U_1 < 0$ and $\Delta U' = U_2 - U_1 < 0$, where jumps of energy values have magnitudes of common order.

Let's examine absolute values of resultant energy disturbances $\Delta U_1$ and $\Delta U_3$ at the boundaries of second body in points $x_1$ and $x_3$ obtained in vicinity of first body, and presented by equations (25) and (26).

Since values $\Delta U$ and $\Delta U'$ are negative and of common order, it takes place $\Delta U' \leq \Delta U \exp\{-\nu_x \lambda_x (x_1 - x_0)\}$ and $\Delta U' \leq \Delta U \exp\{-\nu_x \lambda_x (x_3 - x_0)\}$. Then it follows from equations (25) and (26),

$$\Delta U_1 = -\Delta U' + \Delta U \exp\{-\nu_x \lambda_x (x_1 - x_0)\} \tag{39}$$

$$\Delta U_3 = -\Delta U' + \Delta U \exp\{-\nu_x \lambda_x (x_3 - x_0)\} \tag{40}$$

and from equation (24) for resultant force,

$$\hat{F}(x_1, t) + \hat{F}(x_3, t) = -\nu_x \lambda_x \, k \, \Delta U \left( \exp\{-\nu_x \lambda_x (x_1 - x_0)\} - \exp\{-\nu_x \lambda_x (x_3 - x_0)\} \right) \tag{41}$$

Since $\Delta U < 0$ and $x_3 > x_1 > x_0$ resultant force $\hat{F}(x_1, t) + \hat{F}(x_3, t) > 0$ i.e. force is directed to right way of second body, and second body is repulsed from first body.

Secondly, according to equation (38) we have an estimate of the absolute value of resultant force between two three-dimensional bodies. To get formula for electrostatic forces we look at factor $(\Delta U \cdot V)$ from equation (38). It represents electrostatic energy of solid body but that energy is traditionally expressed via electrostatic charge $Q$. Therefore it follows the Coulomb's formula from equation (38) for repulsive force between two solid bodies that both have negative electrostatic charges $Q = \Delta U \cdot V$ (notice that electrostatic charge $Q$ is concentrated on the body's surface),

$$\left| \hat{F}(t) \right| \cong K \frac{Q_1 Q_2}{R^2} \tag{42}$$

where $Q_1$ and $Q_2$ are electrostatic charges of first and second bodies respectively, $R$ is distance between first and second bodies, and $K > 0$ is Coulomb's constant.

**Figure 5** illustrates repulsive forces between two solid bodies where each body has a negative electrostatic charge.

***b) Attractive Electrostatic Forces***

At first, I consider here one-dimensional space with two solid bodies where first body has negative jump of energy value $\Delta U = U_0 - U_1 < 0$ and second body has positive jump of energy value $\Delta U' = U_2 - U_1 > 0$, and jumps of energy values have magnitudes of common order.

Let's look again at absolute values of resultant energy disturbances $\Delta U_1$ and $\Delta U_3$ at the boundaries of second body in points $x_1$ and $x_3$ obtained in vicinity of first body, and presented by equations (25) and (26).

Since value $\Delta U$ is negative and value $\Delta U'$ is positive and both are of common order, it takes place $\Delta U' \geq \Delta U \exp\{-\nu_x \lambda_x (x_1 - x_0)\}$ and $\Delta U' \geq \Delta U \exp\{-\nu_x \lambda_x (x_3 - x_0)\}$. Then it follows from equations (25) and (26),

$$\Delta U_1 = \Delta U' - \Delta U \exp\{-\nu_x \lambda_x (x_1 - x_0)\} \tag{43}$$

$$\Delta U_3 = \Delta U' - \Delta U \exp\{-\nu_x \lambda_x (x_3 - x_0)\} \tag{44}$$

and from equation (24) for resultant force,

$$\hat{F}(x_1, t) + \hat{F}(x_3, t) = \nu_x \lambda_x \, k \, \Delta U \left( \exp\{-\nu_x \lambda_x (x_1 - x_0)\} - \exp\{-\nu_x \lambda_x (x_3 - x_0)\} \right) \tag{45}$$

Since $\Delta U < 0$ and $x_3 > x_1 > x_0$ resultant force $\hat{F}(x_1, t) + \hat{F}(x_3, t) < 0$ i.e. force is directed to left way of second body, and second body is attracted to first body.

According to equation (38) we have an estimate of the absolute value of resultant force between two three-dimensional bodies. It is represented by the same formula (42).

**Figure 6** illustrates attractive forces between two solid bodies where first body has a negative electrostatic charge, and second body has a positive electrostatic charge.

## 7. Gravitational Forces

This section is devoted to Newton's formula for gravitational forces. We consider two solid bodies where first solid body produces energy disturbance that is much bigger than energy disturbance produced by another solid body. We look at two different cases – situation with rotation of small body around its axis and situation with *tidally locked* small body (i.e. without rotation of small body around axis).

***a) Gravitational Force with Tidally Locked Attraction***

At first, I consider two solid bodies in one-dimensional space with positive jumps of energy values $\Delta U = U_0 - U_1 > 0$ and $\Delta U' = U_2 - U_1 > 0$, when jump of energy value for first body is much bigger than jump of energy value for second body i.e. $\Delta U >> \Delta U'$.

Let's examine the absolute values of resultant energy disturbances $\Delta U_1$ and $\Delta U_3$ at the boundaries of second body in points $x_1$ and $x_3$ obtained in vicinity of first body, and presented by equations (25) and (26).

Since values $\Delta U$ and $\Delta U'$ are both positive and $\Delta U >> \Delta U'$, it takes place $\Delta U' \le \Delta U \exp\{-\nu_x \lambda_x (x_1 - x_0)\}$ and $\Delta U' \le \Delta U \exp\{-\nu_x \lambda_x (x_3 - x_0)\}$. Then it follows from equations (25) and (26),

$$\Delta U_1 = -\Delta U' + \Delta U \exp\{-\nu_x \lambda_x (x_1 - x_0)\} \tag{46}$$

$$\Delta U_3 = -\Delta U' + \Delta U \exp\{-\nu_x \lambda_x (x_3 - x_0)\} \tag{47}$$

and from equation (24) for resultant force,

$$\hat{F}(x_1,t) + \hat{F}(x_3,t) = -\nu_x \lambda_x\, k\, \Delta U \left(\exp\{-\nu_x \lambda_x (x_1 - x_0)\} - \exp\{-\nu_x \lambda_x (x_3 - x_0)\}\right) \tag{48}$$

Since $\Delta U > 0$ and $x_3 > x_1 > x_0$ resultant force $\hat{F}(x_1,t) + \hat{F}(x_3,t) < 0$ i.e. force is pointed to the left way of second body, and second small body is attracted to first massive body.

If we examine results of equations (46) and (47) we see that absolute values $\Delta U_1$ and $\Delta U_3$ are in reality the propagated energy disturbances $-E_1(x_1,t)$ and $-E_1(x_3,t)$ of reflected image of massive solid body $(-\Delta U)$ revealed at the boundaries of small body, and summed with energy disturbances of small body $E_2(x_1,t)$ and $E_3(x_3,t)$. Thus we may say that gravitational force of attraction between massive body and small body is result of interaction of massive body with its reflected image (I suppose from *symmetry* that force, which attracts massive body to small body, is result of interaction of small body with its reflected image as well).

Ultimately according to equation (38) we have an estimate of the absolute value of resultant force between two three-dimensional bodies. To get formula for gravitational forces we examine factor $(\Delta U \cdot V)$ from equation (38). It represents gravitational energy of solid body. I assume (this assumption is in agreement with general assertion that body's energy is proportional to its mass) that energy value $\Delta U$ of solid body is proportional to its density $\rho$ (where $\gamma > 0$),

$$\Delta U = \gamma \rho \tag{49}$$

Merger of equations (38), (49) gives formula for resultant gravitational force $\hat{F}(t)$,

$$\left|\hat{F}(t)\right| \sim \frac{(\rho_1 V_1)(\rho_2 V_2)}{R^2} = \frac{m_1 m_2}{R^2} \tag{50}$$

and ultimately,

$$\hat{F}(t) \cong G \frac{m_1 m_2}{R^2} \tag{51}$$

where $m_1$ and $m_2$ are masses of first and second bodies respectively, $R$ is distance between first and second bodies, and $G$ is gravitational constant.

**Figure 7** illustrates tidally locked gravitational forces between two solid bodies where one body has much bigger jump of energy value than another body does.

***b) Gravitational Force with Rotation of Small Body***

Let us move small body away from massive body in right direction. Then at some moment we have next resultant energy disturbances at the boundaries of small body in points $x_1'$ and $x_3'$ (where $x_3' > x_1' > x_0$),

$$\hat{E}(x_1', t) = -\Delta U' + \Delta U \exp\{-\nu_x \lambda_x (x_1' - x_0)\} > 0 \tag{52}$$

$$\hat{E}(x_3', t) = -\Delta U' + \Delta U \exp\{-\nu_x \lambda_x (x_3' - x_0)\} = 0 \tag{53}$$

At this position force at right boundary of small body in point $x_3'$ becomes zero,

$$\hat{F}(x_1, t) = -\nu_x \lambda_x k \Delta U_1 < 0 \tag{54}$$

$$\hat{F}(x_3,t) = \nu_x \lambda_x k \, \Delta U_3 = 0 \qquad (55)$$

where $\Delta U_1 = \left|\hat{E}(x_1,t)\right|$ and $\Delta U_3 = \left|\hat{E}(x_3,t)\right|$.

If we again move small body in right direction from points $x_1'$ and $x_3'$ on minor distance $\delta > 0$, then resultant energy disturbances are,

$$\hat{E}(x_1' + \delta, t) = -\Delta U' + \Delta U \exp\{-\nu_x \lambda_x (x_1' + \delta - x_0)\} - \varepsilon_1 > 0 \qquad (56)$$

$$\hat{E}(x_3' + \delta, t) = -\Delta U' + \Delta U \exp\{-\nu_x \lambda_x (x_3' + \delta - x_0)\} - \varepsilon_3 < 0 \qquad (57)$$

where $\varepsilon_1 > 0$ and $\varepsilon_3 > 0$ are small.

Therefore resultant energy disturbances at the boundaries of small body in points $x_1' + \delta$ and $x_3' + \delta$ have opposite signs. Contrary boundaries of small body are attracted to each other, and it generates slow rotation of small body. Since $\varepsilon_1 > 0$ and $\varepsilon_3 > 0$ are small, gravitational force still can be approximated by formula (51).

When we move small body further away from massive body, contrary boundaries of small body become stronger attracted to each other and rotation of small body becomes faster. At the same time magnitude of gravitational force acting between small and massive bodies approaches and finally reaches zero. At that moment absolute values of resultant energy disturbances at the boundaries of small body in points $x_1''$ and $x_3''$ become equal,

$$\hat{E}(x_1'', t) = -\Delta U' + \Delta U \exp\{-\nu_x \lambda_x (x_1'' - x_0)\} > 0 \qquad (58)$$

$$\hat{E}(x_3'', t) = -\Delta U' + \Delta U \exp\{-\nu_x \lambda_x (x_3'' - x_0)\} < 0 \qquad (59)$$

$$\hat{E}(x_1'', t) = -\hat{E}(x_3'', t) \quad (60)$$

At this position (when small body is in points $x_1''$ and $x_3''$) the resultant force $\hat{F}(t)$ between small and massive bodies becomes zero.

When we move small body further away from points $x_1''$ and $x_3''$, rotation of small body becomes slow. Gravitational force acting between small and massive bodies changes its direction, and small and massive bodies start being repulsed from each other.

**Figure 8** illustrates situation with gravitational force between small and massive bodies at position when the force equals zero.

## 8. Conclusion

Unfortunately model still doesn't provide answers for all questions. For example, I am trying to understand process behind Newton's third law for forces acting on a distance between two solid bodies.

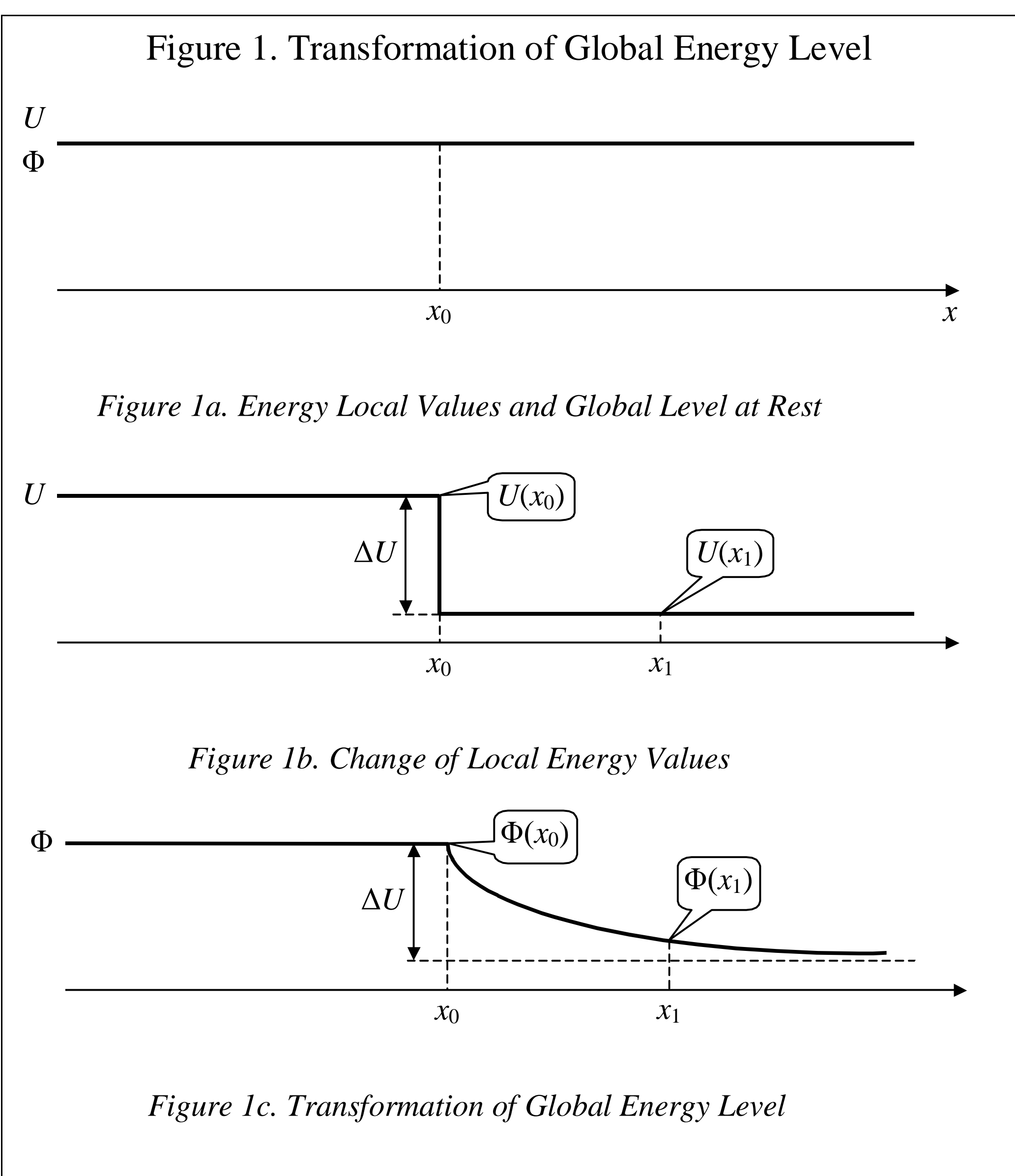

Figure 1. Transformation of Global Energy Level

*Figure 1a. Energy Local Values and Global Level at Rest*

*Figure 1b. Change of Local Energy Values*

*Figure 1c. Transformation of Global Energy Level*

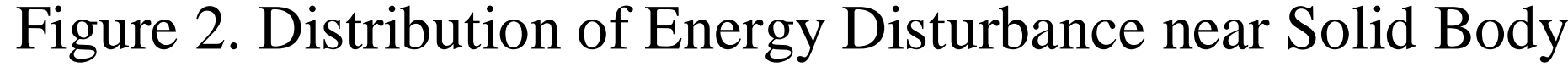

Figure 2. Distribution of Energy Disturbance near Solid Body

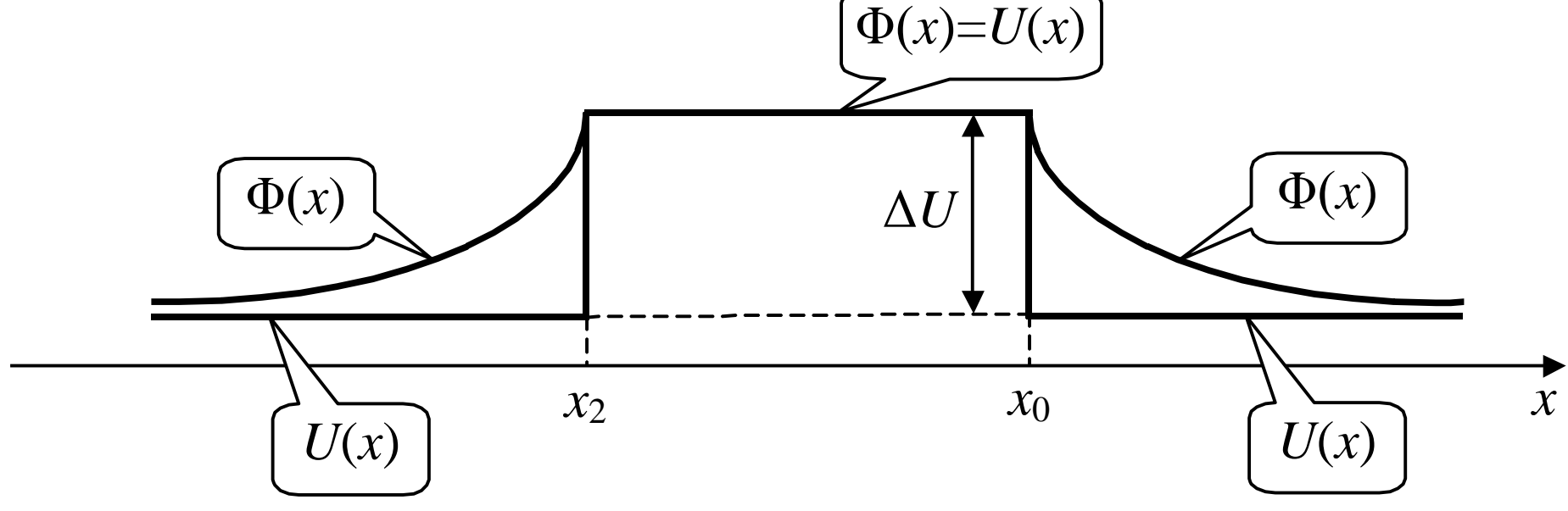

*Figure 2a. Energy Local Values and Global Level near Solid Body*

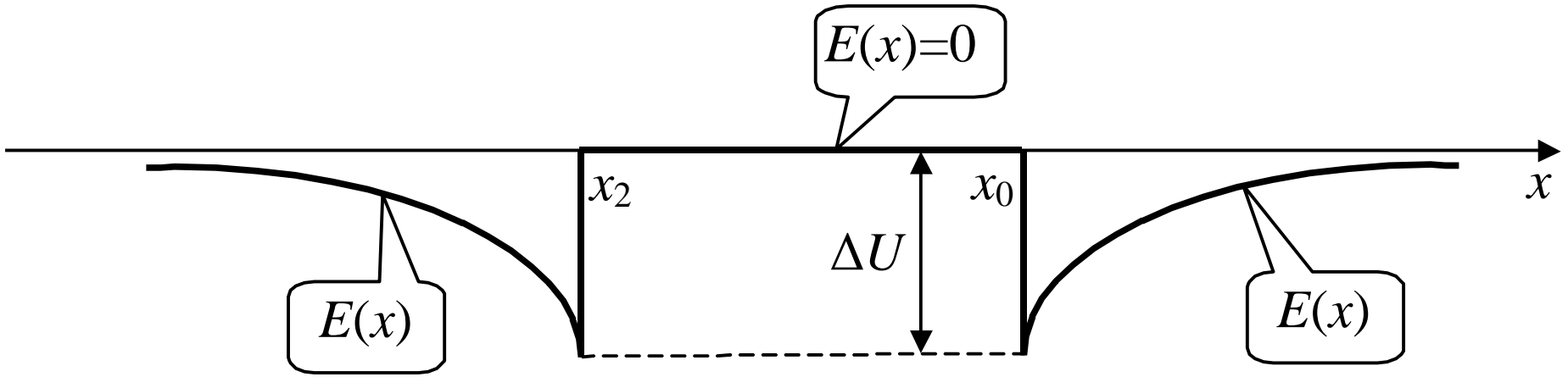

*Figure 2b. Distribution of Energy Disturbance near Solid Body*

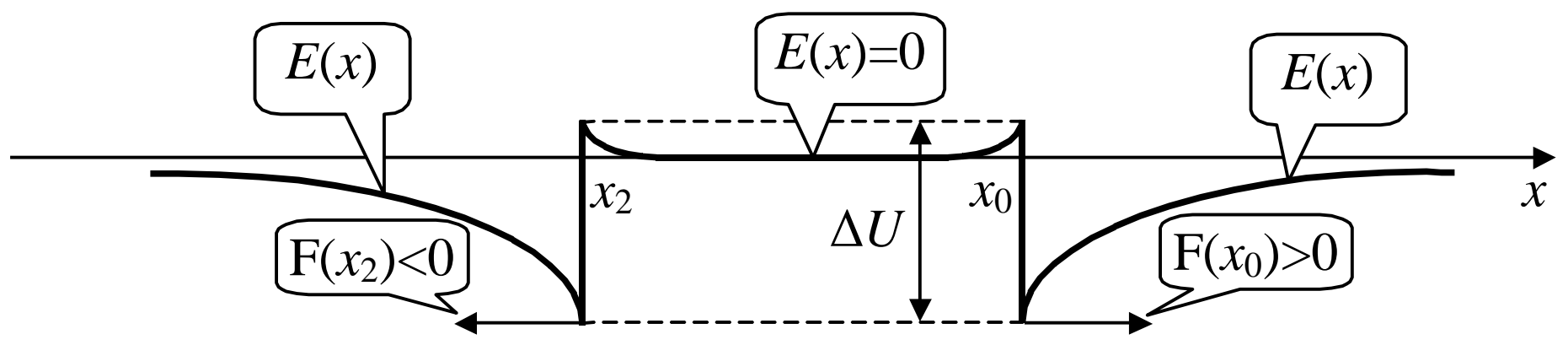

*Figure 2c. Adjusted Distribution of Energy Disturbance*

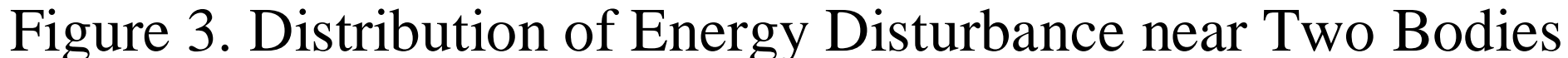

Figure 3. Distribution of Energy Disturbance near Two Bodies

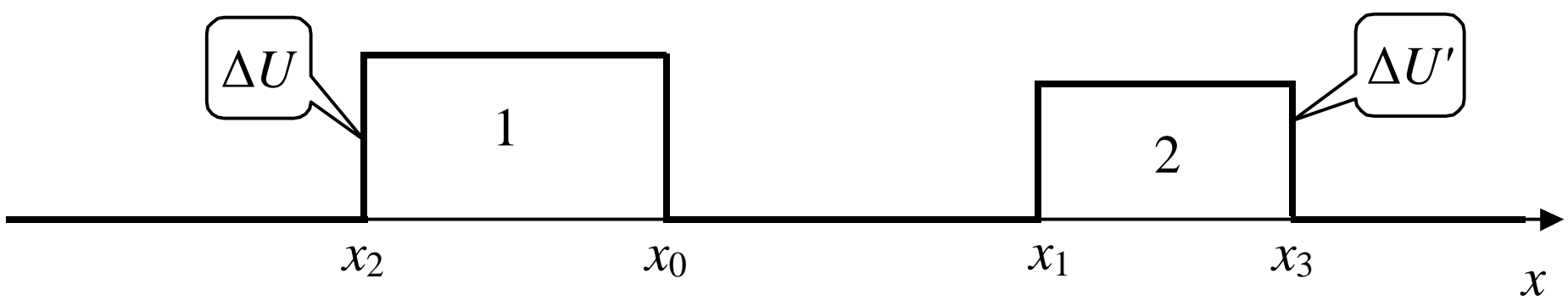

*Figure 3a. Two Solid Bodies in Space and Local Energy Values*

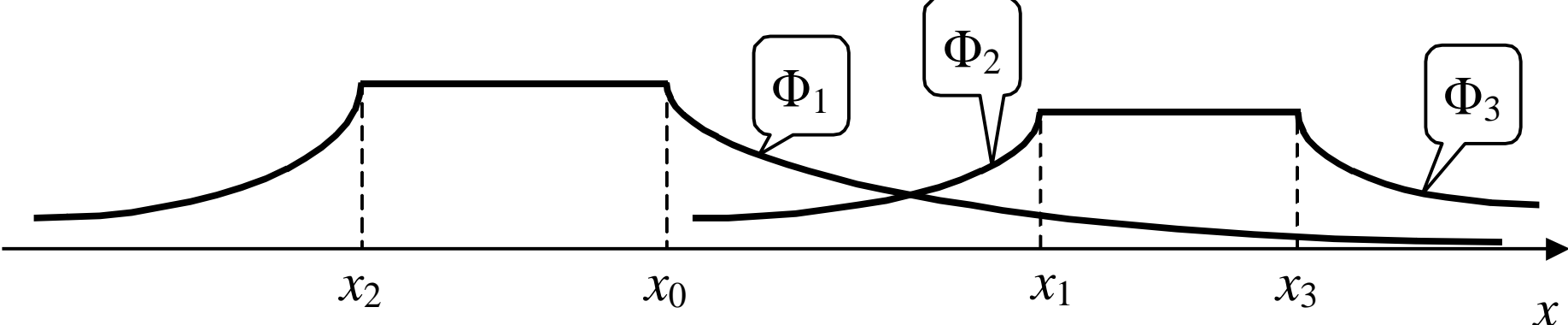

*Figure 3b. Transformation of Global Energy Level*

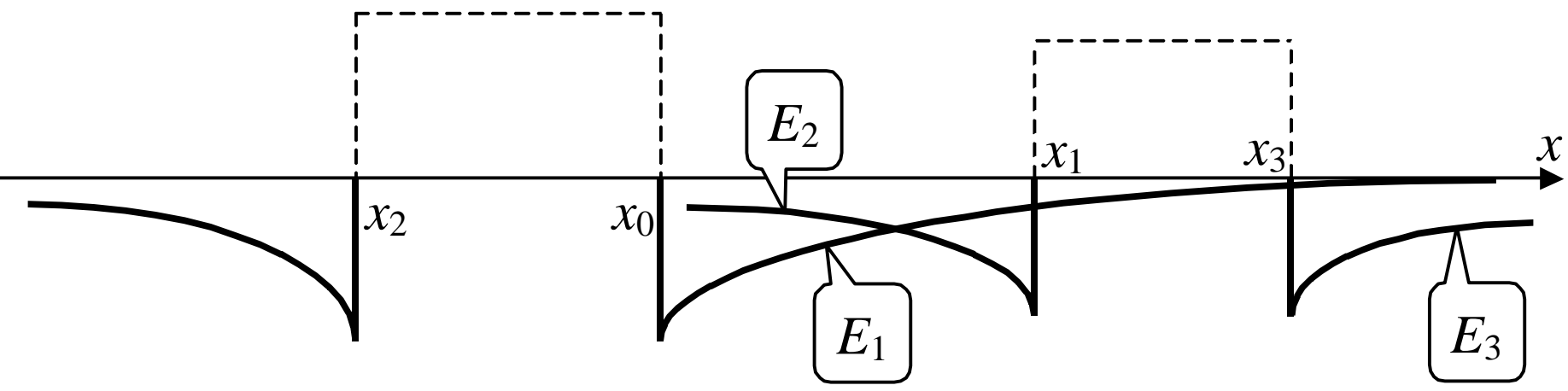

*Figure 3c. Distribution of Energy Disturbances*

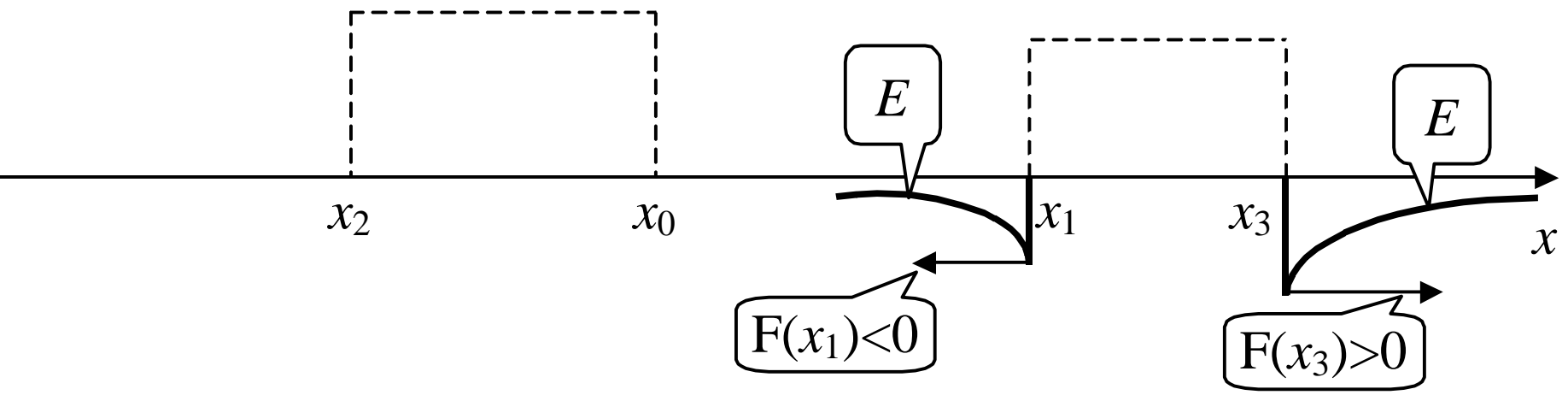

*Figure 3d. Distribution of Resultant Energy Disturbance near Second Body*

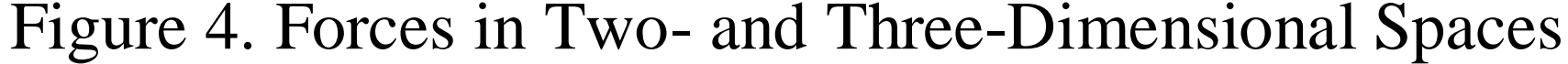

Figure 4. Forces in Two- and Three-Dimensional Spaces

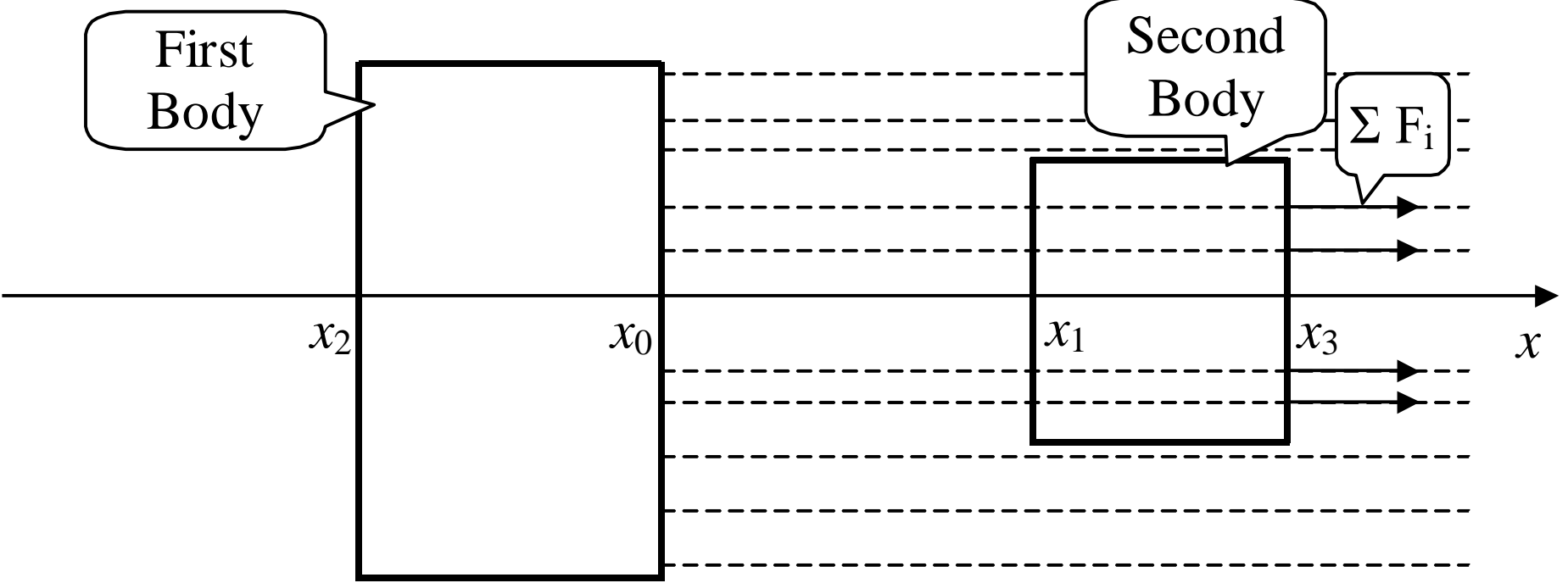

*Figure 4a. Forces Acting on Second Body in Two-Dimensional Space*

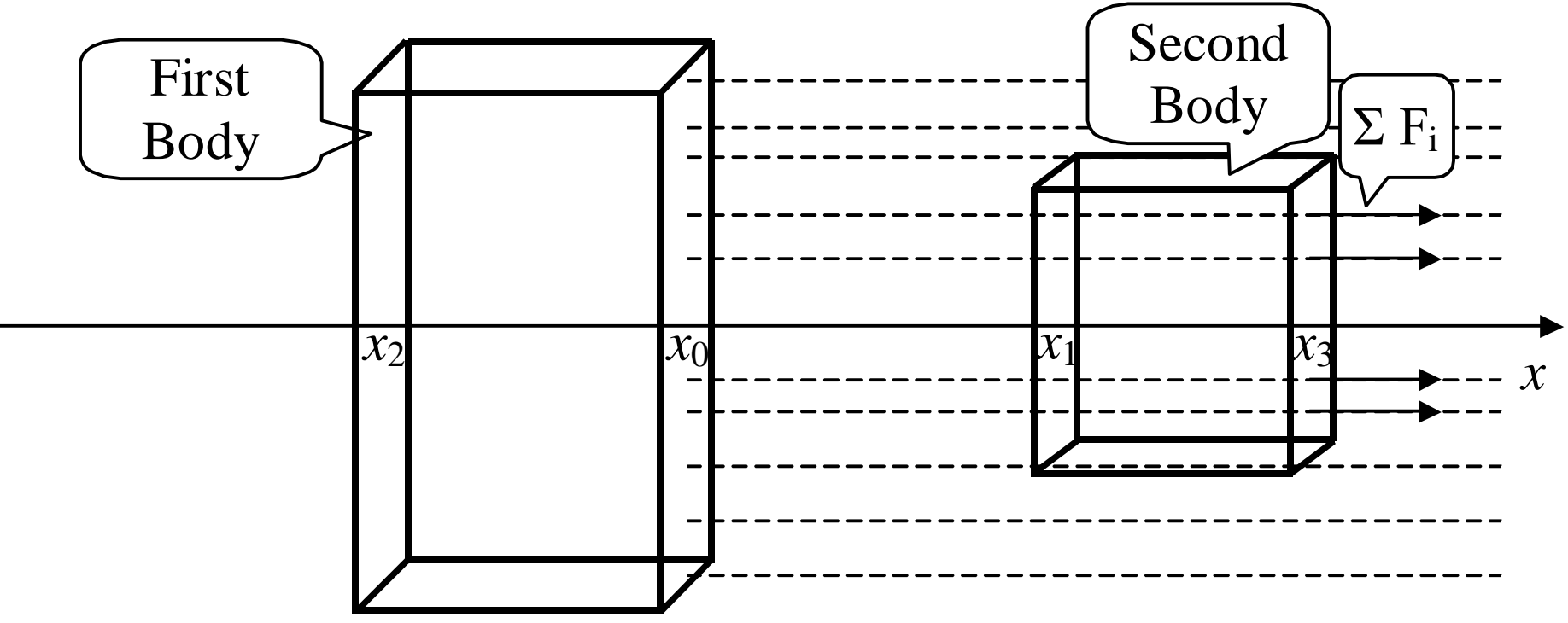

*Figure 4b. Forces Acting on Second Body in Three-Dimensional Space*

Figure 5. Repulsive Electrostatic Forces between Two Bodies

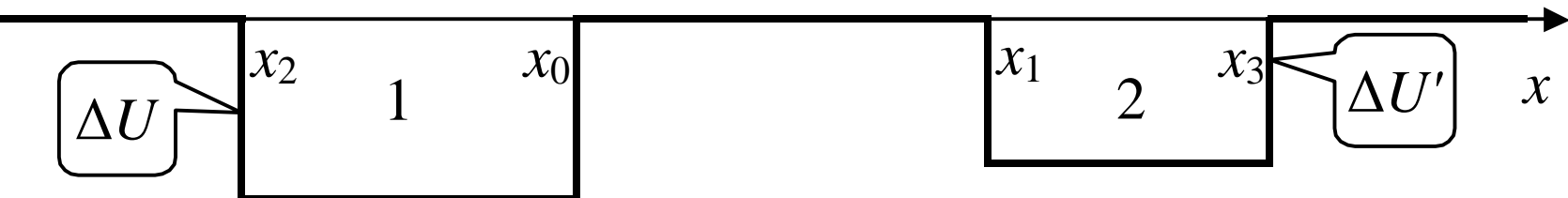

*Figure 5a. Two Solid Bodies with Negative Energy Values in Space*

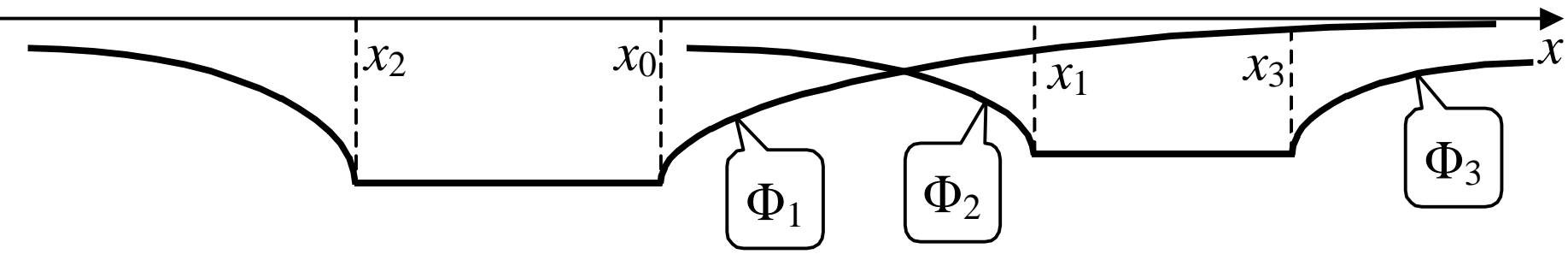

*Figure 5b. Transformation of Global Energy Level*

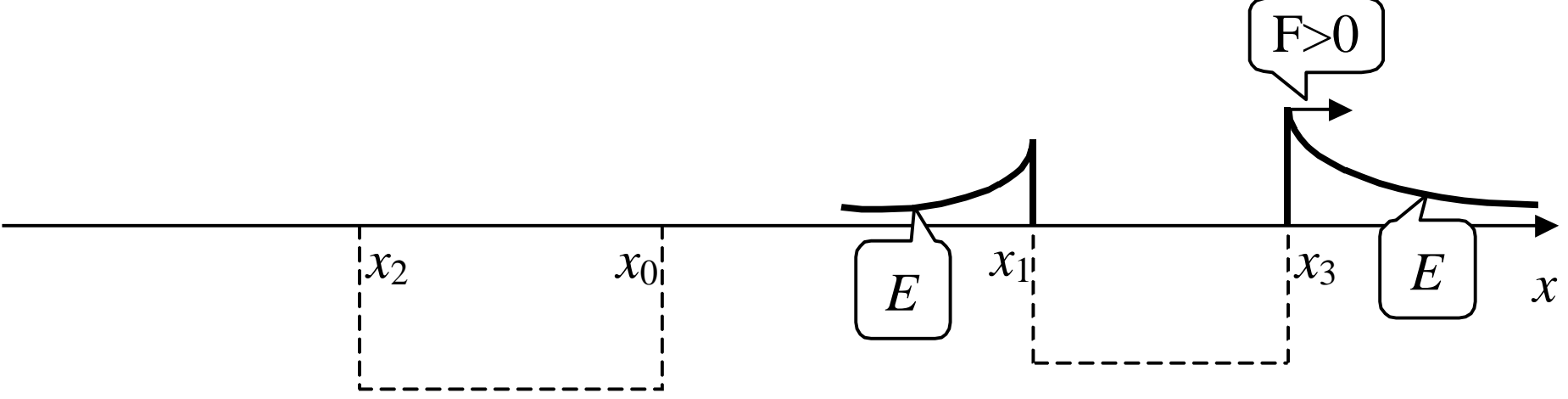

*Figure 5c. Energy Disturbance near Second Body and Repulsive Force*

Figure 6. Attractive Electrostatic Forces between Two Bodies

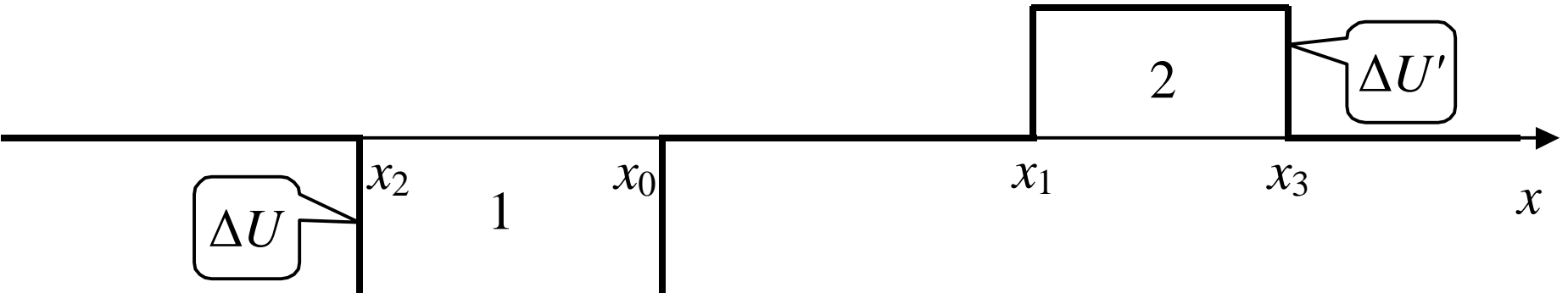

*Figure 6a. Two Solid Bodies with Negative and Positive Energy Values in Space*

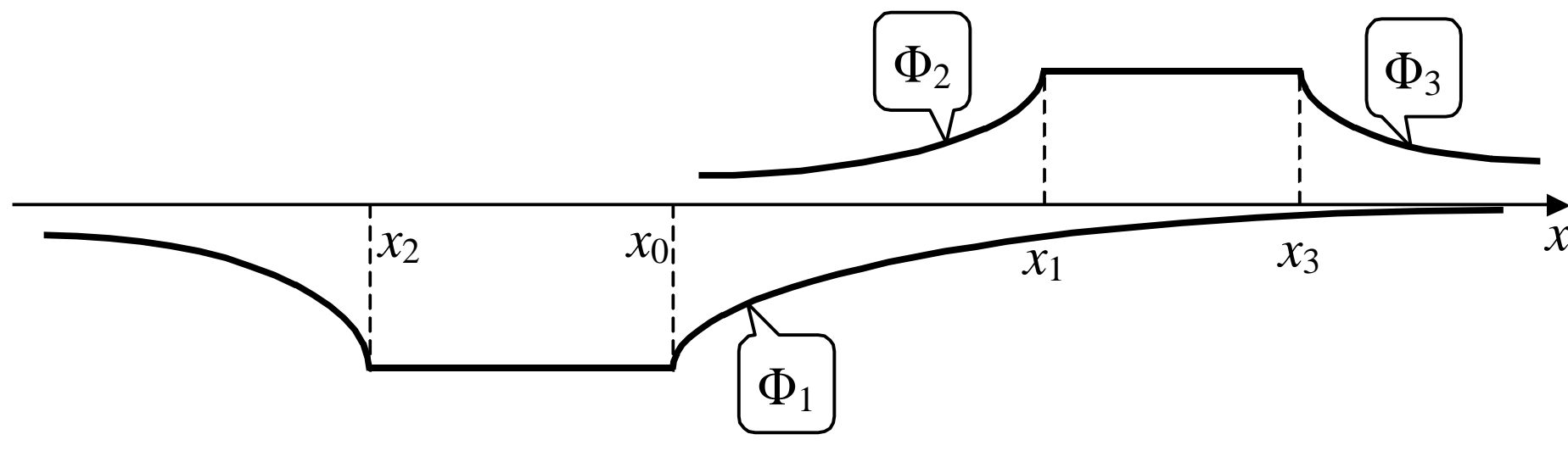

*Figure 6b. Transformation of Global Energy Level*

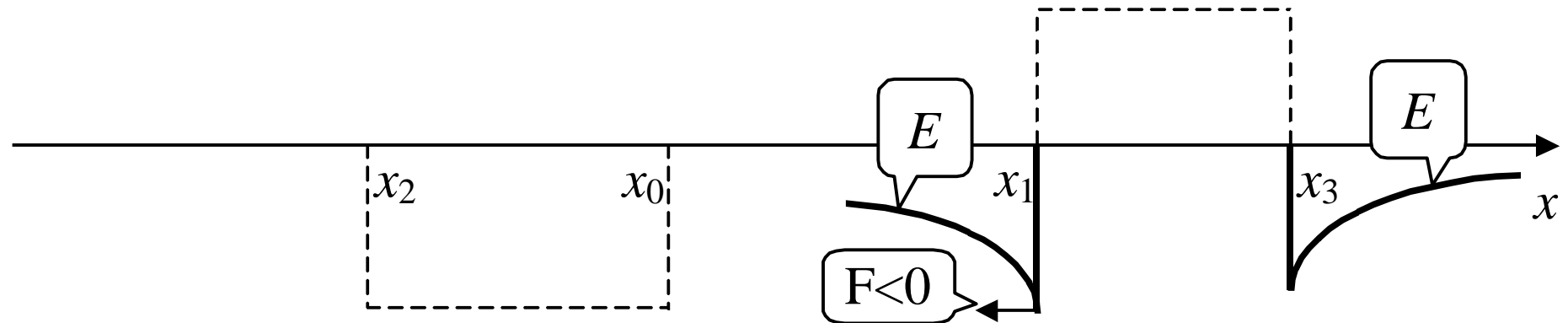

*Figure 6c. Energy Disturbance near Second Body and Attractive Force*

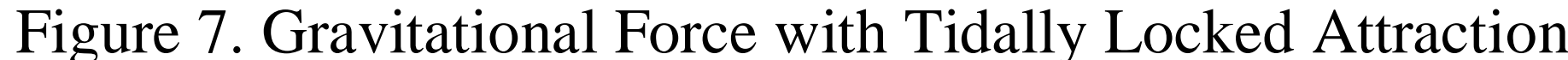

Figure 7. Gravitational Force with Tidally Locked Attraction

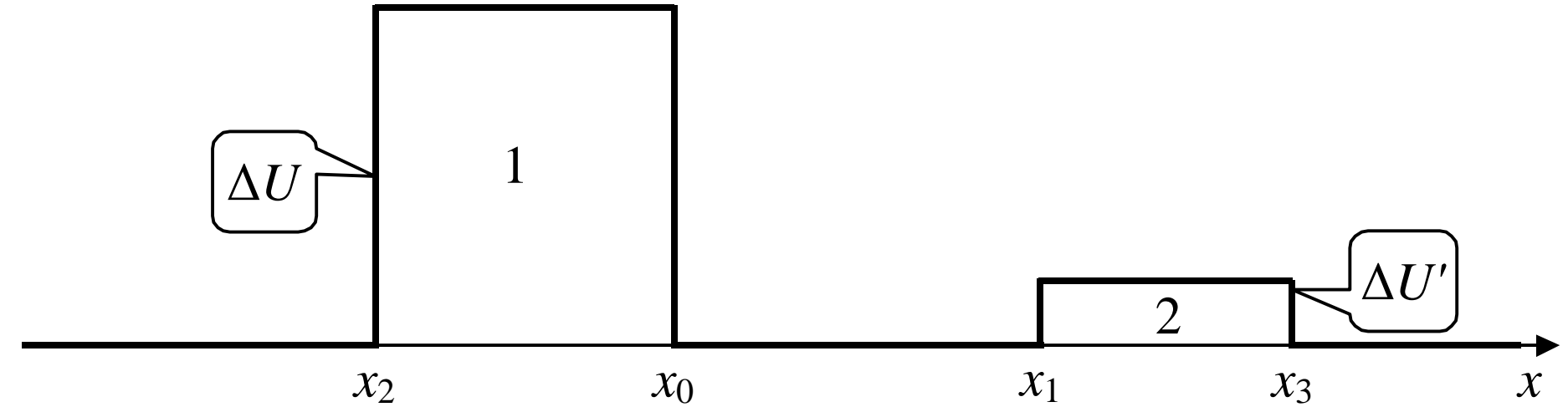

*Figure 7a. Massive and Small Solid Bodies with Positive Energy Values in Space*

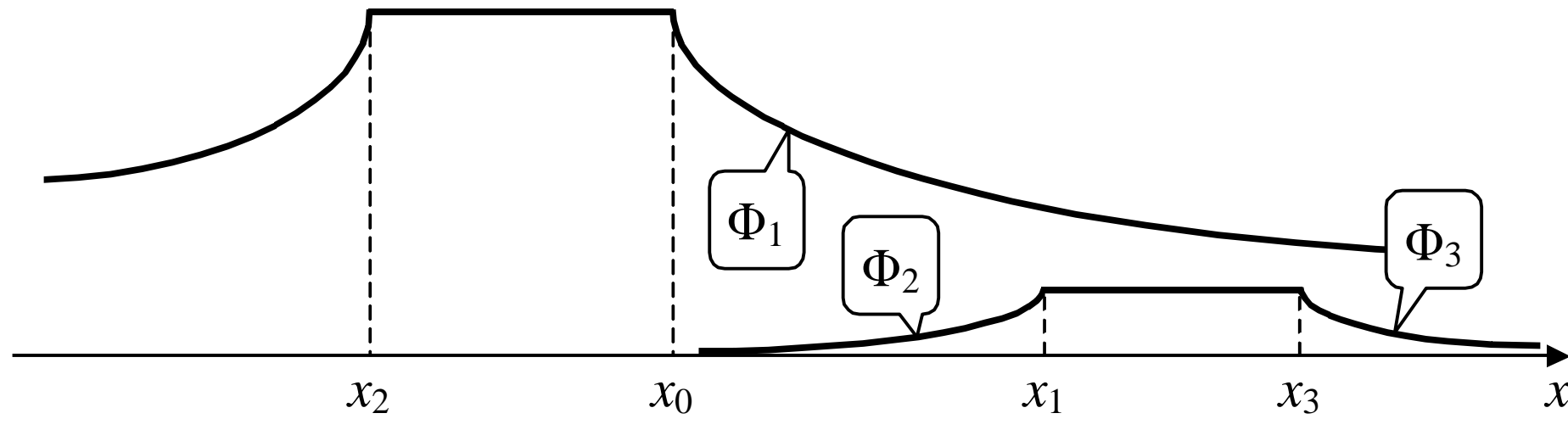

*Figure 7b. Transformation of Global Energy Level*

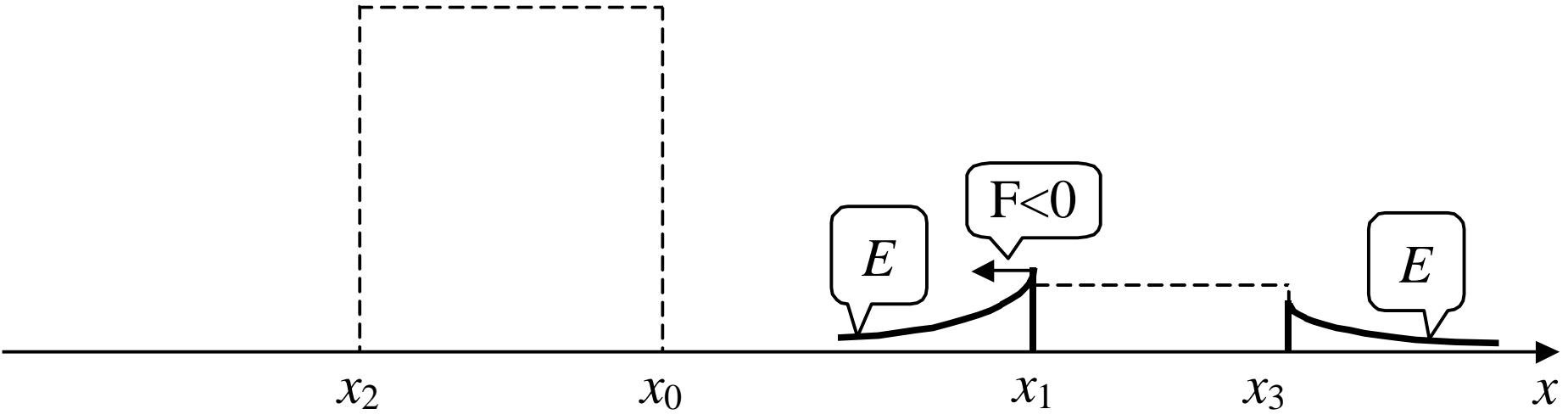

*Figure 7c. Energy Disturbance near Second Body and Gravitational Force*

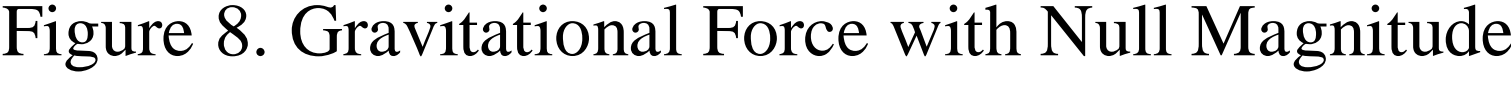

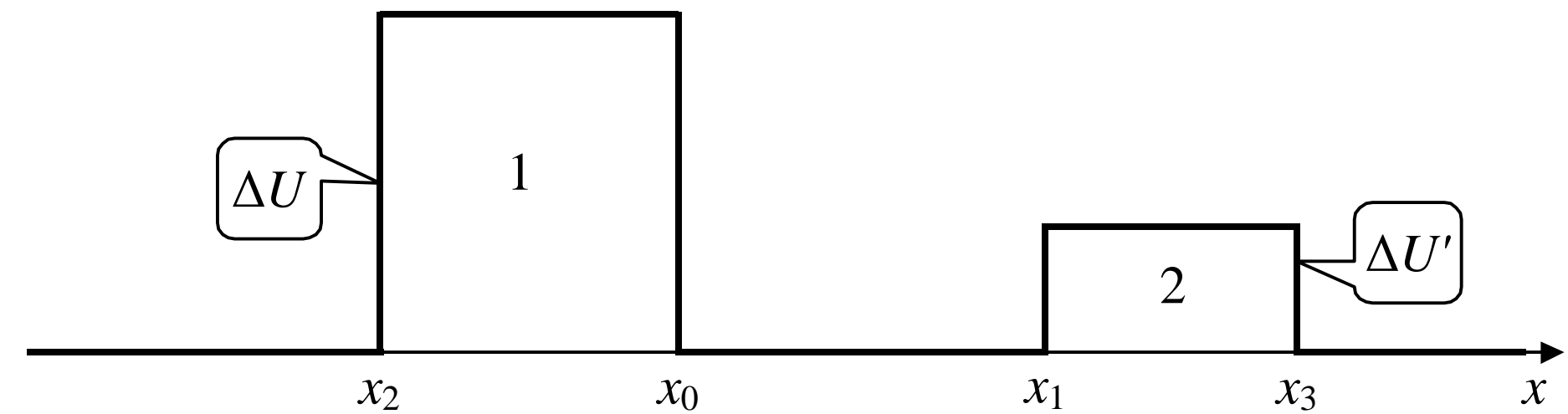

*Figure 8a. Massive and Small Solid Bodies with Positive Energy Values in Space*

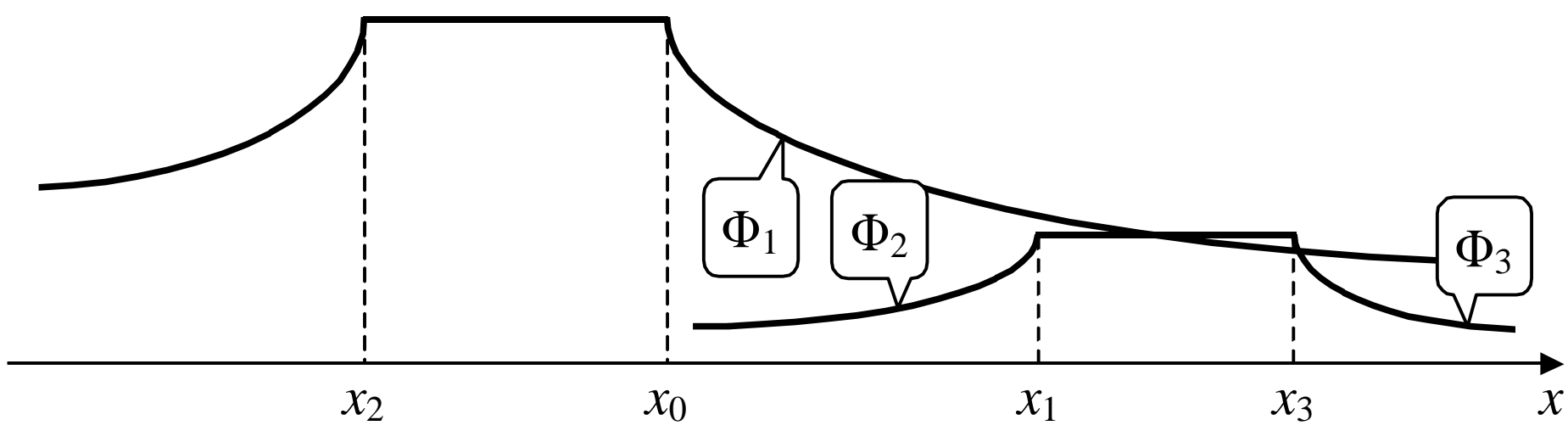

*Figure 8b. Transformation of Global Energy Level*

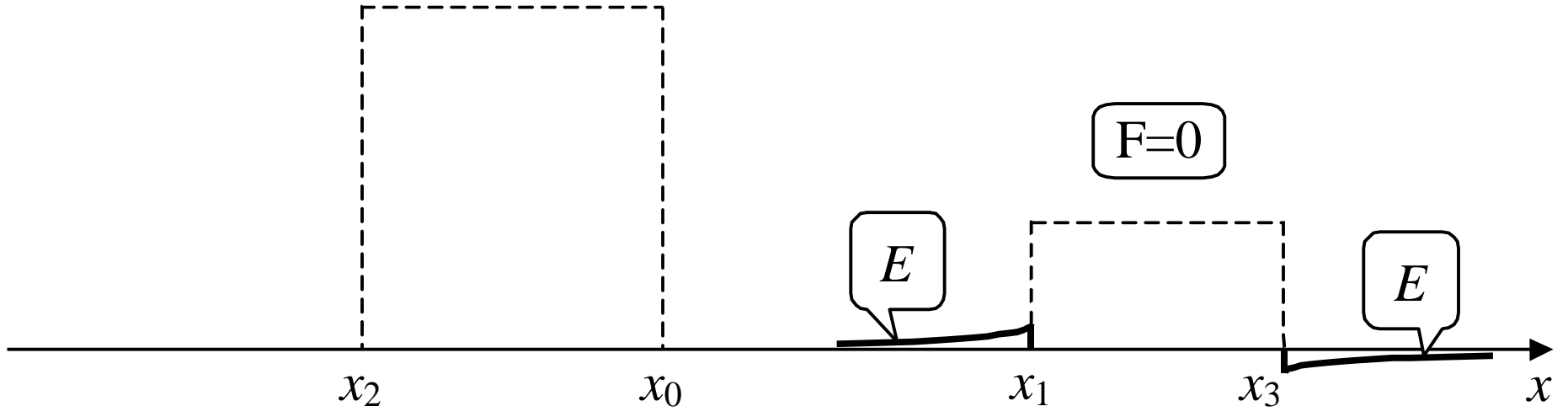

*Figure 8c. Energy Disturbance near Second Body and Null Gravitational Force*